\documentclass[twocolumn,amsmath,amssymb,superscriptaddress]{revtex4}

\usepackage{graphicx} 
\usepackage{dcolumn}  
\usepackage{bm}       

\begin{document}

\title{Electron correlations and bond-length fluctuations in
copper oxide superconductors: electron versus hole doping }

\author{L. Hozoi}
\affiliation{Max-Planck-Institut f\"{u}r Festk\"{o}rperforschung,
             Heisenbergstrasse 1, 70569 Stuttgart, Germany }

\author{S. Nishimoto}
\affiliation{Max-Planck-Institut f\"{u}r Physik Complexer Systeme,
               N\"{o}thnitzer Str. 38, 01187 Dresden, Germany}

\date{\today}

\begin{abstract}
We investigate the nature of the electronic ground state and
electron--lattice couplings for doped chains of CuO$_4$ plaquettes
or CuO$_6$ octahedra.
The undoped configuration implies here Cu $3d^9$ and O $2p^6$
formal valence states.
The results of multiconfiguration calculations on 4-plaquette
(or 4-octahedra) linear clusters indicate strong electron--lattice
interactions and polaronic behavior of the doped particles,
for both electron and hole doping.
For certain phases of the oxygen-ion half-breathing distortions
a multi-well energy landscape is predicted.
Since each well is associated to carriers localized at different
sites, the half-breathing displacements induce charge transfer along
the chain.
In the case of hole-doping, the trends found by \textit{ab initio}
multiconfiguration calculations on 4-octahedra clusters are confirmed
by density-matrix renormalization-group calculations for a $p\!-\!d$,
extended Hubbard model with chains of few tens of CuO$_4$ plaquettes.
Under the assumption of charge separation and the formation of
1/3-doped stripes, our results seem to support the traveling
charge-density wave scenario proposed in some recent contributions
for superconductivity  in copper oxides.
\end{abstract}


\maketitle

\section{Introduction}

A remarkable property of the copper oxide compounds is that
superconductivity can be obtained in these systems by both
hole and electron doping.
Holes added to the insulating, antiferromagnetic (AFM) CuO$_2$
network have O $2p$ character, see for example
\cite{cuo_stollhoff_91,cuo_stollhoff_98,cuo_RLmartin93,
cuo_RLmartin96,cuo_calzado_00,cuo_HNY}.
Results of recent \textit{ab initio}, multiconfiguration
calculations on clusters including several CuO$_6$ octahedra also
indicate strong electron\,--\,lattice couplings \cite{cuo_HNY}.
It was found in Ref.\,\cite{cuo_HNY} that the lowest energy
configurations imply Zhang--Rice (ZR) \cite{ZR_86} like,
quasi-localized states on distorted CuO$_4$ plaquettes.
However, O-atom half-breathing displacements that restore the
high-symmetry structure with identical Cu--O distances induce strong
charge redistribution.
The O $2p$ hole can be (partially) transferred onto a single anion
to give an electronic wave-function with a dominant contribution
from a ...--\,Cu\,$d^9$--\,O\,$p^5$--\,Cu\,$d^9$--... configuration.
The ZR like singlet can hop thus within the CuO$_2$ plane via
such $d^9$--\,$p^5$--\,$d^9$ states through coupling to the oxygen
vibrations. According to the results from \cite{cuo_HNY}, the
energy barrier associated with this double-well potential is
300--400\,meV for an isolated O $2p$ hole.

For not too high doping levels, these ZR type singlets are
sufficiently far apart such that they do not overlap.
It seems that at the optimal hole concentration $p\!\approx\!1/6$
and low temperatures, spin correlations and hole--hole Coulomb
interactions determine within the CuO$_2$ plane an unique, charge
inhomogeneous arrangement of the O $2p$ holes, see for example
\cite{cuo_SM_92,cuo_SM_94,kivelson_stripes03,CuO_bishop03_tu,jbg_03}.
J. B. Goodenough \cite{jbg_03} argued that at $p\!\approx\!1/6$
the doped holes form alternating, hole-free and 1/3-doped stripes
of CuO$_4$ plaquettes.
Using the \textit{ab initio} results from Ref.\,\cite{cuo_HNY}
for hole-doped clusters, the charge and spin distribution for a
1/3-doped stripe could be schematically represented as in Fig.\,1(a).
The squares correspond to ZR states on CuO$_4$ plaquettes with
shorter Cu--O bonds along the chain \cite{cuo_HNY}.
In the undistorted structure, a quasilocalized O $2p$ hole favors
ferromagnetic couplings between adjacent Cu ions, as initially 
predicted in \cite{cuo_hizhny_88,cuo_zaanen_88,cuo_emery_88}.
A three-site, $d^9$--\,$p^5$--\,$d^9$ spin-polaron model was used
by Kochelaev \textit{et al.} to explain the electron paramagnetic
resonance (EPR) signal at different doping levels in
La$_{2-x}$Sr$_{x}$CuO$_4$ \cite{CuO_Kochelaev97}.

\begin{figure}[b]
\includegraphics[angle=270,width=1.00\columnwidth]{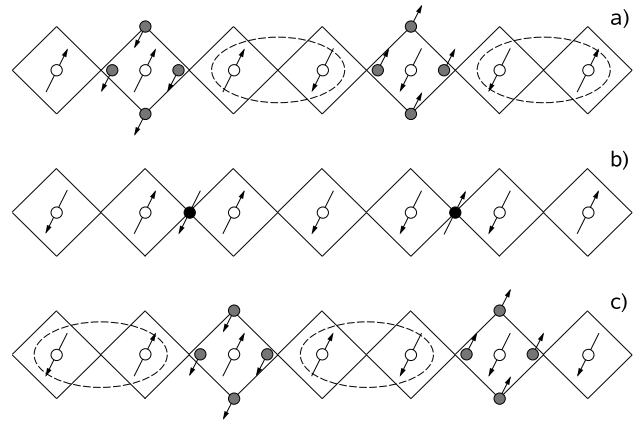}
\caption{ Possible representation of the traveling
charge-density/spin-density wave along a 1/3 hole-doped
CuO$_4$--CuO$_4$--... chain, see Section III and
Ref.\,\cite{cuo_HNY}.
The O $2p$ holes are shown schematically with black or gray color.
The ellipses indicate spin-singlet couplings.}
\end{figure}

We speculated in \cite{cuo_HNY} that for a distribution of the
oxygen holes as that shown in Fig.\,1(a), the strong coupling\
between the nuclear and electronic motions, the hole--hole Coulomb
repulsion, and the in-plane spin interactions would cause strong
correlations among half-breathing O-atom vibrations on every
third, ZR like CuO$_4$ plaquette along the doped chains.
The superconductivity might be related to the collective, coherent
motion of the O $2p$ holes along the 1/3-doped chains, which is
associated with phase coherence among the kind of O-atom
vibrations shown in Fig.\,1(a--c).
This evokes the vibronic mechanism and the traveling
charge-density/spin-density wave (CDW/SDW) scenario suggested by
Goodenough \cite{jbg_03}.

\begin{figure}[b]
\includegraphics[angle=0,width=0.65\columnwidth]{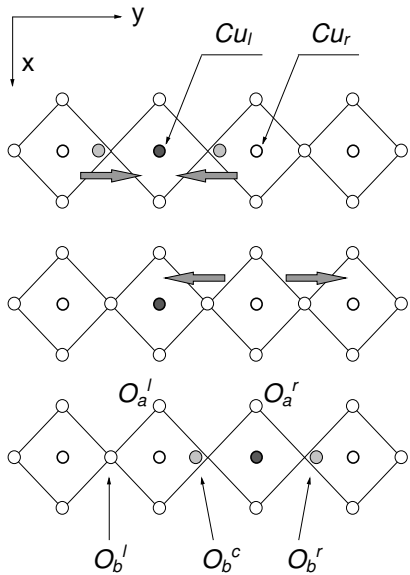}
\caption{ Electron-doped, 4-plaquette cluster and the phases of
the O-ion half-breathing distortions investigated by CAS\,SCF
calculations.
A Cu $3d_{x^2-y^2}^{\,2}$ ion is shown with a filled, dark-gray
circle. From left to right, the transition metal sites are labeled
as Cu$_{ll}$, Cu$_{l}$, Cu$_{r}$, and Cu$_{rr}$.
Ligands that are displaced from the middle positions are shown with
light-gray color.}
\end{figure}

Spatially localized bound states of two electrons were analyzed
within a two-dimensional Holstein--Hubbard model on a square
lattice by Proville and Aubry \cite{QS_99_00}.
Two electrons bound into a localized singlet state as a result
of strong electron--lattice interactions are often referred to
as a bipolaron.
It was found in Ref.\cite{QS_99_00} that for certain values of
the model parameters a ``quadrisinglet'' (QS) state similar to
the ZR like state discussed above can be formed. 
The near-degeneracy between the QS ``bipolaron'' and  a two-site
spin-singlet bipolaron can lead to a sharp reduction of the 
bipolaron effective mass.
Proville and Aubry argued that such near-degeneracy effects plus
quantum tunneling effects associated with a rather flat energy
landscape should be crucial ingredients in the theory of
``bipolaronic'' superconductivity \cite{alexandrov_95_03}.

The fact that the \textit{ab initio} calculations from
Ref.\cite{cuo_HNY} confirm a number of distinct effects
and concepts (that is, ZR type states on CuO$_4$ plaquettes
\cite{ZR_86}, Kondo like interactions and ferromagnetic couplings
between $3d_{x^2-y^2}$ electrons adjacent to a quasilocalized
oxygen hole \cite{cuo_hizhny_88,cuo_zaanen_88,cuo_emery_88},
near-degeneracy between the QS configuration and a two-site
spin-singlet configuration and strong electron--lattice couplings
\cite{QS_99_00}) deduced or proposed on the basis of model
Hamiltonians that rest on quite different approximations and
describe rather different types of interactions is very
interesting.  
It illustrates both the power of and the need for first principles,
explicitly correlated methods in the study of these materials.
Such calculations on not very large clusters are sometimes
essential for understanding the ``local'' many-body physics and
for setting up the ``right'' model for the extended system, as
recently shown in the case of a mixed-valence vanadium oxide,
NaV$_2$O$_5$ \cite{NaVO_phi4}.
Near-degeneracy and pseudo Jahn--Teller effects involving O-hole
$d^1$--\,$p^5$--\,$d^1$ configurations similar to the 
$d^9$--\,$p^5$--\,$d^9$ three-site spin-polaron in the doped
copper oxides appear to be the source of most of the intriguing 
properties of this compound.


In the present paper we report results of \textit{ab initio}
wave-function based, embedded cluster calculations for the
electron-doped case. We investigate the ground-state electron
distribution and electron--lattice couplings.
For an 1-doped-electron, 4-plaquette linear cluster, our data show
that the lowest energy state corresponds to a configuration where
the extra electron is localized on a CuO$_4$ plaquette with
\textit{elongated} Cu--O bonds.
As in the hole-doped system, the O-atom half-breathing
displacements interact with the electronic degrees of freedom and
may induce charge transfer along the CuO chain.
We shall describe various aspects of this kind of interactions and
discuss the similarities and differences with the hole-doped
situation.

\section{1-doped-electron, 4-plaquette linear clusters:
bond-length fluctuations and inter-site charge transfer}

Multiconfiguration, complete active space (CAS)
self-consistent-field (SCF) electronic structure calculations
were carried out on linear clusters including four CuO$_4$ 
plaquettes. The CAS\,SCF method is designed to provide an unbiased
description of systems where near-degeneracy occurs among 
several electron configurations \cite{book_qc}. It is based on a 
partitioning of the total orbital space into three subsets, 
inactive, active, and virtual orbitals. The inactive orbitals
are doubly occupied in all configurations and should be chosen
as the orbitals that are not expected to contribute to 
near-degeneracy correlation effects. The active orbitals are  
subject to no restriction on their occupation numbers and the
virtual orbitals are unoccupied in all configurations.
The CAS\,SCF wave-function is constructed as a full 
configuration-interaction (CI) expansion in the configurational
space spanned by the active orbitals. Since both the CI coefficients 
and the orbitals are variationally optimized, the CAS\,SCF 
wave-function is extremely flexible.
It is this flexibility that makes the CAS\,SCF method well suited 
for the description of competing valence structures, bond breakings,
and dissociation processes.   

We used the crystal parameters reported for the electron-doped copper
oxide superconductor compound Sr$_{1-y}$Nd$_{y}$CuO$_2$
\cite{jbg_SrNdCuO2_91}, with $y\!\approx\!0.16$, Cu--O distances
of 1.97\,\AA , and no apical oxygens.
The 4-plaquette cluster was embedded in an array of point charges
that reproduce the Madelung potential associated with an idealized
SrCuO$_2$ ionic system.
We imposed $C_{2v}$ symmetry restrictions, with $xy$ and $yz$
mirror planes (see Fig.2), and employed the following gaussian-type,
atomic natural orbital basis sets for the [Cu$_4$O$_{13}$] cluster:
Cu $21s15p10d6f/5s4p3d1f$ and O $14s9p4d/4s3p1d$ \cite{ANOs_CuO}.
It is known by now that such basis sets provide a quite 
accurate description of both ground-state and excited-state
properties in a large variety of transition metal (TM) oxides,
see for example
\cite{cuo_calzado_00,NaVO_phi4,cuo_munoz_02,cuo_calzado_02} 
and \cite{coen_thesis}. 
The Cu and Sr neighboring ions were represented by total ion
potentials \cite{TIPs_CuSr}. The calculations were performed with
the \textsc{molcas} program package \cite{molcas6}.

With one electron added to the undoped, formally Cu\,$d^9$\,O\,$p^6$
configuration, the minimal active orbital space should consist of
four (or even three) open-shell $3d_{x^2-y^2}$ orbitals.
However, trial CAS\,SCF calculations were carried out with few
virtual orbitals added to the active space. It turned out that
these extra orbitals have either Cu $4s$ or Cu $3d'$ character,
where $3d'$ is a more diffuse orbital than $3d$ \cite{dd_77_81},
and non-negligible occupation numbers.
Also, large effects were observed on the absolute energies, as
compared to the minimal active space.

The relevance of the $4s$ orbital to the local-density approximation
(LDA) conduction bands and to the in-plane/inter-plane hoppings was
analyzed for several copper oxide compounds in
Refs.\cite{oka_95,oka_01}.
A clear correlation was observed \cite{oka_01} between the value of
the critical temperature $T_c$ and the energy and specific
composition of the so-called axial orbital, a hybrid with mixed
Cu $4s$, Cu $3d_{3z^2-r^2}$, and apical-oxygen $2p_z$ character.
The importance of the Cu $4s$ and $4p$ orbitals for an accurate
description of the electron occupation of the Cu $3d_{x^2-y^2}$
and O $2p_x$/$2p_y$ orbitals in model Hamiltonian calculations
and also their role in screening the on-site Cu $3d$ Coulomb
repulsion was discussed in \cite{cuo_stollhoff_98}.

We investigated the dependence of the cluster total energy on
various distortions and varied the positions of the Cu ions and
of those oxygens (O$_b$, see Fig.2) bridging the cations.
For such a cluster, energetic effects associated with displacements
of the O$_a$ oxygen ions, see Fig.2, are not meaningful, because
part of the nearest neighbors of these ligands are modeled by 
effective potentials or even point charges.
Nevertheless, under the assumption of charge segregation into
filamentary stripe segments and one-dimensional conduction
\cite{CuO_bishop03_tu,jbg_03,cuo_hizhny_88}, the distortions along
the $y$ axis should be the most relevant. 
We found that the cluster energy is minimized for a geometry
where the Cu--O$_b$ bonds on one of the CuO$_4$ plaquettes are
stretched by approximately 8\% and the nearest Cu neighbors are
also displaced by about 1\% of the initial Cu--O bond length, each
of them in the same direction as the adjacent oxygen.
The doped electron is accommodated in the $3d_{x^2-y^2}$ orbital
on the plaquette with elongated Cu--O$_b$ bonds to give a Cu
$3d_{x^2-y^2}^{\,2}$ configuration.
We report results for geometries where either the Cu$_l$--O$_b$
or Cu$_r$--O$_b$ bonds are stretched, see Fig.2, and discuss the
changes in the charge distribution for these two plaquettes and
the variation of the cluster energy.
However, in this study we keep all Cu ions at fixed positions. This
has only minor effects on the ground-state wave-function and energy.
Our active space includes four $3d_{x^2-y^2}$, the Cu$_l$
$3d_{3z^2-r^2}$ and Cu$_r$ $3d_{3z^2-r^2}$ orbitals, and four
virtuals.
This is a 9\,electrons/10\,orbitals CAS, with about fifty thousand
determinants.

For the minimum-energy state with stretched Cu$_l$--O$_b$ bonds
and a Cu$_l$ $3d_{x^2-y^2}^{\,2}$ electron configuration, for
example, those four correlating orbitals taken from the virtual
orbital space in order to enlarge the minimal CAS have Cu$_l$
$3d'_{x^2-y^2}$, Cu$_l$ $4s$, Cu$_r$ $3d'_{3z^2-r^2}$ and Cu$_r$
$3d'_{x^2-y^2}$ character. There is some mixing between the Cu$_l$
$3d_{3z^2-r^2}$ and $4s$ basis functions. The occupation of the
Cu$_l$ $3d_{3z^2-r^2}$ -like orbital is $1.97$ and the
predominantly Cu$_l$ $4s$ orbital has an occupation number of
$0.03$. There are also large contributions of the Cu $4s$
functions to some of the other active orbitals. The Cu$_r$
$3d_{3z^2-r^2}$ and $3d'_{3z^2-r^2}$ orbitals have occupation
numbers of $1.99$ and $0.01$, respectively. The occupation numbers
of the Cu$_l$ $3d'_{x^2-y^2}$ and Cu$_r$ $3d'_{x^2-y^2}$ orbitals
are somewhat smaller than $0.01$.

Inclusion of the diffuse $d'$ orbitals into the active space is
necessary for describing changes in the radial charge distribution
for different $d^n$ configurations in the multiconfigurational 
wave-function.
For free TM atoms (ions), this is a correlation effect involving
pairs of electrons in the $4s^23d^n$, $4s3d^{n+1}$, and $3d^{n+2}$
states \cite{dd_77_81}.
Occupation numbers as large as 0.1 for the $d'$ orbitals of the 
Ni atom \cite{dd_92} demonstrate that a careful analysis is
needed when dealing with TM atoms or TM compounds.
In the copper oxide systems, the different $d^n$ configurations
may be related to both on-site Cu $3d$--$4s$ and inter-site, 
Cu\,--\,Cu and O\,--\,Cu, excitations. 
Radial, or ``breathing'', correlation effects may also occur for 
the O $2p$ orbitals. When the ligand to metal charge transfer
excitations are relevant \cite{ppdd_coen_01}, a highly accurate
CAS\,SCF wave-function would require the inclusion of two extra
sets of orbitals into the active space, ligand $2p$ orbitals 
involved in the O\,$2p$\,--\,Cu\,$3d$ bonding plus the 
corresponding $2p'$ orbitals \cite{coen_thesis,ppdd_coen_01}.
Such an extension of the active space is not possible for our
cluster, due to the high computational cost. Nevertheless, the
low occupation numbers of the $3d'_{x^2-y^2}$ orbitals, see
the preceding paragraph, suggest that the essential
near-degeneracy correlation effects are already included in
our model. 
It will be shown in the next section that the role of the O $2p'$
orbitals becomes important in the hole-doped case. 


\begin{figure}[!t]
\includegraphics[angle=0,width=1.00\columnwidth]{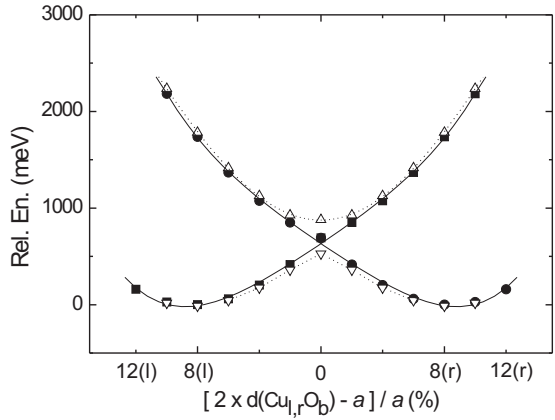}
\caption{ Results of static, total energy calculations for an
electron-doped, 4-plaquette cluster, see text and Fig.2.
The filled symbols are the CAS\,SCF data.
The lowest points correspond to Cu$_l$ $3d_{x^2-y^2}^{\,2}$ or
Cu$_r$ $3d_{x^2-y^2}^{\,2}$ configurations on plaquettes with
Cu--O$_b$ distances larger by $8\%$.
For geometries where the Cu--O$_b$ bonds on the \textit{adjacent}
plaquette are stretched by more than $12\%$, an 1-electron
Cu$_l$\,$d_{x^2-y^2}$--\,Cu$_r$\,$d_{x^2-y^2}$ charge transfer
occurs.
The dotted curves are obtained by CAS\,SI calculations.
The CAS\,SCF energies for the structures with CuO$_b$ bonds
stretched by $8\%$ on one of the Cu$_l$O$_4$ or Cu$_r$O$_4$
plaquettes are taken as reference.
$a$ is the in-plane lattice constant.}
\end{figure}

The potential energy landscape when stretching the
Cu$_l$--O$_b$ or Cu$_r$--O$_b$ bonds is shown in Fig.3\,. The
middle point on the horizontal axis corresponds to the undistorted
cluster. From each minimum-energy configuration, the two O$_b$
ions on the Cu $3d_{x^2-y^2}^{\,2}$ plaquette are shifted back to the
undistorted structure in steps of $2\%$ of the high-symmetry Cu--O
distance. The orbitals from the previous step are used each time
as starting orbitals. Once the undistorted geometry is reached we
start stretching the Cu--O$_b$ bonds on the \textit{adjacent}
plaquette, see Fig.2.
Each curve in Fig.3 is obtained by following this procedure. It turns
out that the initial electronic wave-function remains largely
unchanged for a relatively broad range of Cu--O$_b$ distortions.
Its character changes abruptly when the Cu--O$_b$ bonds on the
adjacent plaquette are elongated by about $12\%$. Only for such a
geometry the doped electron is transferred to the other
$d_{x^2-y^2}$ orbital.

\begin{table*}[!ht]
\caption{Mulliken charges per centre and basis function type for
the minimum-energy Cu$_l$ $d_{x^2-y^2}^{\,2}$ configuration on an
electron-doped, 4-plaquette linear cluster. The Cu$_l$\,--\,O$_b$
bonds are stretched by $8\%$. Values for the Cu$_l$O$_4$ and
Cu$_r$O$_4$ plaquettes are listed. The Mulliken populations on the
other plaquettes, Cu$_{ll}$O$_4$ and Cu$_{rr}$O$_4$, are very
similar to those on Cu$_r$O$_4$.
The charge of each $3d$ $t_{2g}$ atomic orbital is 2.0\,.
There is also some charge associated with the Cu $f$ and O $d$
functions, not shown in the table.}
\begin{ruledtabular}
\begin{tabular}{lrrrrrrr}
                     &O$_b^l$  &O$_a^l$  &Cu$_l$    &O$_b^c$
                     &Cu$_r$   &O$_a^r$  &O$_b^r$            \\
\colrule
O $2s$, \ \,Cu $4s$  &1.68     &1.83     &0.60      &1.68
                     &0.67     &1.82     &1.70               \\
O $2p_x$, Cu $4p_x$  &1.99     &1.96     &0.14      &1.99
                     &0.22     &1.89     &1.99               \\
O $2p_y$, Cu $4p_y$  &1.75     &2.00     &0.09      &1.74
                     &0.23     &2.00     &1.70               \\
O $2p_z$, Cu $4p_z$  &1.87     &1.97     &0.08      &1.87
                     &0.17     &1.96     &1.89               \\
Cu $d_{3z^2-r^2}$    &         &         &1.83      &
                     &1.96     &         &                   \\
Cu $d_{x^2-y^2}$     &         &         &\textbf{2.00}  &
                     &1.18     &         &                   \\
Total el. charge     &9.30     &9.77     &28.78     &9.28
                     &28.54    &9.67     &9.29               \\
\end{tabular}
\end{ruledtabular}
\end{table*}

\begin{table*}[!ht]
\caption{Mulliken charges per centre and basis function type for
a 6-plaquette linear cluster with two doped electrons.
The Cu$_{ll}$\,--\,O$_b$ and Cu$_{rr}$\,--\,O$_b$ bonds are stretched
by $8\%$.
Values for Cu and O$_b$ ions on the four central plaquettes,
Cu$_{ll}$O$_4$, Cu$_l$O$_4$, Cu$_r$O$_4$, and Cu$_{rr}$O$_4$, are
listed.
The same notations as in Fig.2. are used.}
\begin{ruledtabular}
\begin{tabular}{lrrrrrrrrrr}
                    &O$_b^{ll}$ &Cu$_{ll}$    &O$_b^l$ &Cu$_l$
                    &O$_b^c$    &Cu$_r$       &O$_b^r$ &Cu$_{ll}$
                    &O$_b^{rr}$ \\
\colrule
O $2s$, \ \,Cu $4s$ &1.68       &0.58         &1.67    &0.67
                    &1.69       &0.67         &1.67    &0.58
                    &1.68\\
O $2p_x$, Cu $4p_x$ &1.99       &0.12         &1.99    &0.22
                    &1.99       &0.22         &1.99    &0.12
                    &1.99\\
O $2p_y$, Cu $4p_y$ &1.73       &0.09         &1.74    &0.24
                    &1.72       &0.24         &1.74    &0.09
                    &1.73\\
O $2p_z$, Cu $4p_z$ &1.86       &0.08         &1.87    &0.17
                    &1.89       &0.17         &1.87    &0.08
                    &1.86\\
Cu $d_{3z^2-r^2}$   &           &1.82         &        &1.97
                    &          &1.97         &        &1.82
                    &    \\
Cu $d_{x^2-y^2}$    &           &\textbf{2.00}&        &1.18
                    &           &1.18         &        &\textbf{2.00}
                    &    \\
Total el. charge    &9.26       &28.73        &9.27    &28.55
                    &9.30       &28.55        &9.27    &28.73
                    &9.26\\
\end{tabular}
\end{ruledtabular}
\end{table*}

We list in Table\,I Mulliken populations of atomic orbitals on the
Cu$_l$O$_4$ and Cu$_r$O$_4$ plaquettes for the Cu$_l$
$d_{x^2-y^2}^{\,2}$ lowest energy configuration. The charges
associated with the O $2s$ and O $2p_x$/$2p_y$ functions
overlapping the Cu $d_{x^2-y^2}$ orbitals are sensibly smaller
than $2.00$, even on the $d_{x^2-y^2}^{\,2}$ plaquette. The
\textquotedblleft missing\textquotedblright charge is in the Cu
$4s$ and $4p$ orbitals. This deviation from the fully ionic
picture was previously pointed out in
Ref.\,\cite{cuo_stollhoff_98}, for example. Between the two
plaquettes, the major differences are related to the populations
of the Cu $3d_{x^2-y^2}$, $3d_{3z^2-r^2}$, $4s$, $4p$, and O$_a$
$2p_x$ orbitals.

If the matrix elements between the Cu$_l$ $d_{x^2-y^2}^{\,2}$ and
Cu$_r$ $d_{x^2-y^2}^{\,2}$ electronic wave-functions are ignored,
the adiabatic potential energy curves for the O-ion half-breathing
displacements discussed above intersect at the point associated
to the undistorted cluster, see Fig.3\,.
However, if the electronic structures are allowed to relax by mixing
of the Cu$_l$ $d_{x^2-y^2}^{\,2}$ and Cu$_r$ $d_{x^2-y^2}^{\,2}$
configurations, no intersection would occur. A so-called avoided
crossing is expected for the potential curves.
The matrix elements between non-orthogonal, separately optimized
CAS\,SCF wave-functions can be obtained by State Interaction (SI)
calculations \cite{SI_86_89}.
The potential energy curves corresponding to the CAS\,SI states
are shown in Fig.3 by dashed lines.
The interaction between the non-orthogonal states is not too large,
with energetic effects of not more than few tenths of eV.
The weights of the non-orthogonal states in the lowest CAS\,SI root
are both 1/2 at the middle point.
For chain bonds stretched by $8\%$ on one of the plaquettes, the
weights are 0.99 and 0.01\,.
The energy barrier associated with the lower CAS\,SI curve is
540\,meV.

For a more accurate estimate of the size of the energy barrier,
polarization effects beyond the boundaries of the linear,
4-plaquette cluster should also be taken into account. 
Test calculations on a 5-plaquette cluster where all four nearest
neighbor plaquettes of a ``central'' Cu$_c$ $d_{x^2-y^2}^{\,2}$
ion are treated at the all-electron level show indeed that such 
effects are important. 
We used for these calculations $D_{2h}$ symmetry restrictions and
the following basis sets \cite{ANOs_CuO}: Cu $21s15p10d6f/5s4p3d2f$
for the five TM ions, O $14s9p4d/4s3p2d$ for the ligands bridging
two cations and O $14s9p/4s3p$ for the other twelve oxygens.
For a Cu$_c$ $d_{x^2-y^2}^{\,2}$ configuration on the plaquette at
the center of the cluster, the CAS\,SCF energy difference between
the undistorted structure and the geometry where the Cu$_c$--O
distances along the $y$ axis are longer by $8\%$ is 130\,meV
lower than for the linear, 4-plaquette cluster.
Applying such a correction to the CAS\,SI results, the energy
barrier would decrease to about 400\,meV. Longer range polarization
effects should further lower this value.

We also performed calculations on a 6-plaquette linear cluster with
two doped electrons.
The following contractions were applied for the Cu and O basis sets:
Cu $21s15p10d/5s4p3d$ and O $14s9p/4s3p$ \cite{ANOs_CuO}.
Ground-state Mulliken populations for the geometry with Cu--O$_b$
bonds stretched by $8\%$ on the Cu$_{ll}$O$_4$ and Cu$_{rr}$O$_4$
plaquettes are given in Table\,II.
The results were obtained with an active space including 12
electrons and 12 orbitals.
For this geometry and choice of the multiconfigurational space, the
CAS\,SCF active orbitals were identified as six Cu $3d_{x^2-y^2}$
orbitals, the Cu$_{ll}$ $3d_{3z^2-r^2}$, $4s$ and  Cu$_{rr}$
$3d_{3z^2-r^2}$, $4s$ plus two correlating Cu$_{ll}$ $3d'_{x^2-y^2}$
and Cu$_{rr}$ $3d'_{x^2-y^2}$.
The occupation numbers of these active orbitals are similar to those
found for the smaller cluster. The Mulliken populations are very
similar, too.
The results show thus that localized $3d_{x^2-y^2}^{\,2}$ states
are also obtained in larger clusters and for more than a single
doped electron.
For the electron-doped case, an analysis of inter-carrier couplings
by CAS\,SCF calculations on such larger clusters will be provided in
a forthcoming publication.
In the case of hole-doping, preliminary CAS\,SCF calculations on
2-oxygen-hole, 6-octahedra clusters reveal that the hole--hole
interaction is large. Due to the strong mutual repulsion
within the O $2p$ ``band'', the two holes are pushed to the
boundaries of the cluster and meaningful CAS\,SCF results cannot be
obtained in this situation.
However, inter-carrier correlations for the hole-doped case are
addressed in the next section by a model Hamiltonian approach that
enables calculations on chains of few tens of plaquettes.

The existence of a local double-well potential like that shown in
Fig.3 for Cu--O bond-length distortions in both n-doped and p-doped
layered Cu oxides was predicted by Goodenough \cite{jbg_03}.
Goodenough also suggested that around optimal doping the doped
carriers display some particular (dynamic) ordering within each
layer, perhaps an arrangement with alternating undoped and
1/3-doped chains of plaquettes.
The coupling between these ordered electrons/holes and in-plane O-ion
half-breathing phonons is believed to be at the heart of
high-temperature superconductivity
\cite{CuO_bishop03_tu,jbg_03,cuo_hizhny_88}.
Evidence for stripe formation in electron-doped cuprates was recently
reported in Ref.\cite{CuO_eldo_sun04}, for example.

\section{Hole-doped system}

\begin{table*}[!ht]
\caption{Major contributions of the oxygen $2p$ and metal $3d$ atomic
orbitals to those active (natural) orbitals implying $2p\!-\!2p$
and $2p\!-\!3d$ bonding on the Cu$_l$O$_4$ and Cu$_r$O$_4$
plaquettes.
CAS\,SCF results for a hole-doped, 4-plaquette linear cluster.
Coefficients smaller than 0.20 are not listed.
For each geometry, the nature of the other active orbitals is
discussed in the text.}
\begin{ruledtabular}
\begin{tabular}{lrrrrrrrr}
  State:
&\multicolumn{4}{c}{O$_b^c$-hole, undistorted cluster}
&\multicolumn{4}{c}{Cu$_l$O$_4$ ZR, distorted Cu$_l$--\,O$_b$ bonds
                    \tablenotemark[1]}                            \\
\cline{2-5}
\cline{6-9}
Orb. Occ. No.:        &1.99      &1.96      &1.84      &0.21
                      &1.99      &1.97      &1.90      &0.13      \\
\colrule

O$_b^l$ $p_y$         &          &0.54      &0.22      &0.37
                      &0.29      &0.67      &0.41      &0.42      \\
O$_a^l$ $p_x$ (x2)    &$\pm$0.20 &$\pm$0.25 &$\pm$0.26 &$\pm$0.31
                      &$\mp$0.42 &          &$\pm$0.29 &$\pm$0.29 \\
Cu$_l$ $d_{x^2-y^2}$  &          &          &--0.61    &0.74
                      &          &          &--0.63    &0.77      \\
O$_b^c$ $p_y$         &0.37      &0.51      &--0.56    &--0.47
                      &--0.29    &0.67      &--0.41    &--0.42    \\
Cu$_r$ $d_{x^2-y^2}$  &          &          &          &--0.21
                      &          &          &          &          \\
O$_a^r$ $p_x$ (x2)    &$\mp$0.37 &$\pm$0.21 &          &
                      &          &          &          &          \\
O$_b^r$ $p_y$         &0.39      &          &          &
                      &          &          &          &          \\
\end{tabular}
\end{ruledtabular}
\tablenotetext[1]{ \ The chain Cu$_l$--O$_b$ bonds are 6\% shorter.}
\end{table*}

Multiconfiguration calculations for hole-doped clusters have been
performed in Ref.\,\cite{cuo_HNY}. However, a minimal active space
was applied in that study, with a single orbital for each Cu $3d$
or O $2p$ hole. In this section we report results of CAS\,SCF
calculations with larger active spaces. We use the same
4-octahedra linear cluster and the same (static) embedding as in
Ref.\,\cite{cuo_HNY}, with the lattice parameters of
La$_{1.85}$Sr$_{1.85}$CuO$_4$ \cite{LaCuO_cava87}.
The basis sets are also the same, Cu $21s15p10d/5s4p3d$,
O $14s9p4d/4s3p1d$ for the bridging O$_b$ oxygens, and O
$14s9p/4s3p$ for the other oxygens \cite{ANOs_CuO}.
As already mentioned in Introduction, the CAS\,SCF results indicate
that ZR type singlet states are only formed on CuO$_4$ plaquettes
with Cu--O bonds \textit{shorter} than the average Cu--O distance
deduced from experiment.
On the other hand, for the undistorted geometry the
doped hole has the largest weight onto a single ligand and in the
simplest picture the dominant contribution to the wave-function
comes from a ...--\,Cu\,$d^9$--\,O\,$p^5$--\,Cu\,$d^9$--...
configuration \cite{note_localization}.
The latter state is few hundreds meV higher in
energy than the former. Also for the hole-doped case, O-ion
half-breathing displacements determine thus a
multi-well energy landscape and can induce charge transfer along a
chain of CuO$_4$ plaquettes. We note that geometry optimizations
for transition metal oxides like MnO and NiO at the Hartree--Fock
(HF) level \cite{geopt_mno_nio,geopt_cu2o} slightly overestimate
the interatomic distances. This means that the multi-well energy
landscape would also be obtained with such HF optimized lattice
parameters \cite{note_HFgo}. For the electron-doped case, the
usual overestimation of the lattice constants, $2\!-\!4\%$
\cite{geopt_mno_nio,geopt_cu2o}, is significantly lower than the
range of the Cu--O bond distortions associated with the multi-well
energy curve, $\pm8\%$.

As in the previous section, we added to the minimal active space
orbitals from both the inactive, doubly occupied set and the
virtual space. Trial calculations showed that the most important
near-degeneracy correlation effects are already accounted for with
extensions of only few orbitals. In Tables III, IV, and V we show
results obtained from CAS\,SCF calculations with 13 electrons and
12 orbitals in the active space. Compared to the minimal
(\textit{i.e.}, 5\,electrons/5\,orbitals) CAS, four inactive and
three virtual orbitals were moved into the active set. Depending
on the input starting orbitals, different CAS\,SCF solutions can
be obtained for the undistorted cluster, with the doped hole
having the largest weight onto one of the three bridging oxygens,
O$_b^l$, O$_b^c$ or O$_b^r$. The largest contributions to those
active orbitals implying $2p$\,--\,$2p$ and $3d$\,--\,$2p$ bonding
on the central Cu$_l$O$_4$ and Cu$_r$O$_4$ plaquettes for the
CAS\,SCF solution with the doped hole having the largest weight
onto the O$_b^c$ ligand are illustrated in Table\,III.
We also display in Table\,III the occupation numbers of those
orbitals.
There are other three active orbitals with Cu$_{ll}$, Cu$_r$, and
Cu$_{rr}$ $3d_{x^2-y^2}$ character and occupation numbers of
nearly 1.
For this CAS\,SCF solution, the remaining active orbitals have
O$_b^c$ $2s$, Cu$_l$ $3d_{3z^2-r^2}$, O$_b^c$ $2p'_{y}$,
Cu$_l$\,$4s$\,--\,O$_b^c$\,$2s'$\,--\,Cu$_r$\,$4s$, and Cu$_l$
$3d'_{3z^2-r^2}$ character. Their occupation numbers are 1.999,
1.98, 0.02, 0.01, and 0.01, respectively.

The Mulliken electron
populations for the relevant atomic functions on the Cu$_l$O$_4$
and Cu$_r$O$_4$ plaquettes are listed in Table\,IV. Compared to
the minimal-CAS calculations from Ref.\,\cite{cuo_HNY}, the charge
disproportion between the (O$_b^l$ $2p_y$, O$_a^l$ $2p_x$) and
O$_b^c$ $2p_y$ orbitals is significantly smaller. The difference
between the O$_b^l$ $2p_y$ and O$_b^c$ $2p_y$ Mulliken charges,
for example, is 0.13 instead of 0.50. We also note the
Cu$_l$--\,O$_b^c$--\,Cu$_r$ asymmetry. A symmetry equivalent
solution, with a pair of bonding and antibonding
...--\,O$_b^c$\,$2p_y$--\,Cu$_r$\,$3d_{x^2-y^2}$--... orbitals
that is symmetry equivalent to the pair of bonding and antibonding
...--\,Cu$_l$\,$3d_{x^2-y^2}$--\,O$_b^c$\,$2p_y$--... orbitals in
Table\,III (the third and fourth columns) can also be obtained.
The CAS\,SCF wave-function changes its character from one type to
the other for displacements of the O$_b^c$ ion of $\pm0.8\%$ around
the middle position.
This is very different in comparison with the electron-doped case in
the previous section, where the character of the CAS\,SCF
wave-function remains nearly unchanged, \textit{i.e.} either Cu$_l$
$d_{x^2-y^2}^{\,2}$ or Cu$_r$ $d_{x^2-y^2}^{\,2}$, for distortions
of up to $\pm12\%$ of the Cu--O$_b$ bonds.

\begin{table*}[!ht]
\caption{Mulliken charges per centre and basis function type for a
broken-symmetry oxygen-hole state on a 4-octahedra linear cluster.
The doped hole has the largest weight on the O$_b^c$ ion.
Charges for the central Cu$_l$O$_4$ and Cu$_r$O$_4$ plaquettes are
listed. The Cu--O distances are all the same.}
\begin{ruledtabular}
\begin{tabular}{lccccccc}
                    &O$_b^l$   &O$_a^l$   &Cu$_l$   &O$_b^c$
                    &Cu$_r$    &O$_a^r$   &O$_b^r$            \\
\colrule

O $2s$, \ \,Cu $4s$ &1.59         &1.81         &0.77   &1.58
                    &0.67         &1.81         &1.60        \\
O $2p_x$, Cu $4p_x$ &2.00         &\textbf{1.65}&0.28   &2.00
                    &0.26         &1.81         &2.00        \\
O $2p_y$, Cu $4p_y$ &\textbf{1.65}&2.00         &0.22
                                                &\textbf{1.51}
                    &0.17         &2.00         &1.83        \\
O $2p_z$, Cu $4p_z$ &1.93         &1.97         &0.25   &1.92
                    &0.23         &1.97         &1.93        \\
Cu $d_{3z^2-r^2}$   &             &             &2.00   &
                    &2.00         &             &            \\
Cu $d_{x^2-y^2}$    &             &             &1.06   &
                    &1.15         &             &            \\
Total el. charge    &9.19         &9.42         &28.70  &9.03
                    &28.57        &9.59         &9.37        \\
\end{tabular}
\end{ruledtabular}
\end{table*}

\begin{table*}[!ht]
\caption{Mulliken charges per centre and basis function for a
minimum-energy ZR type configuration on the Cu$_l$O$_4$ plaquette.
CAS\,SCF results for an 1-doped-hole, 4-octahedra linear cluster,
see text. The Cu$_l$\,--\,O$_b$ bonds are shorter by
$6\%$. Values for the central Cu$_l$O$_4$ and Cu$_r$O$_4$
plaquettes are listed.}
\begin{ruledtabular}
\begin{tabular}{lccccccc}
                     &O$_b^l$ &O$_a^l$   &Cu$_l$   &O$_b^c$
                     &Cu$_r$  &O$_a^r$   &O$_b^r$               \\
\colrule

O $2s$, \ \,Cu $4s$  &1.57         &1.81         &0.82   &1.57
                     &0.64         &1.80         &1.60        \\
O $2p_x$, Cu $4p_x$  &2.00         &\textbf{1.65}&0.29   &2.00
                     &0.25         &1.83         &2.00        \\
O $2p_y$, Cu $4p_y$  &\textbf{1.59}&2.00         &0.25
                                                 &\textbf{1.59}
                     &0.15         &2.00         &1.84        \\
O $2p_z$, Cu $4p_z$  &1.92         &1.97         &0.25   &1.92
                     &0.23         &1.97         &1.93        \\
Cu $d_{3z^2-r^2}$    &             &             &2.00   &
                     &2.00         &             &            \\
Cu $d_{x^2-y^2}$     &             &             &1.02   &
                     &1.16         &             &            \\
Total el. charge     &9.10         &9.43         &28.74  &9.09
                     &28.53        &9.61         &9.38        \\
\end{tabular}
\end{ruledtabular}
\end{table*}

With the larger active space, the ZR like state with shorter
Cu--O$_b$ distances ($6\%$ shorter) on one of the Cu$_l$O$_4$ or
Cu$_r$O$_4$ plaquettes is lower in energy by about 460\,meV. Major
contributions to the relevant $2p\!-\!3d$ active orbitals are
shown in Table\,III. The doped hole is now approximately equally
distributed over the O$_b$ $p_y$ and O$_a$ $p_x$ atomic orbitals
on the distorted plaquette, see the pair of $d\!-\!p$ bonding and
antibonding orbitals with lower occupation numbers in Table\,III. As
above, there are three active orbitals with Cu$_{ll}$, Cu$_r$, and
Cu$_{rr}$ $3d_{x^2-y^2}$ character and occupation numbers of
roughly 1. The other active orbitals have Cu$_l$ $3p_y$, Cu$_l$
$3d_{3z^2-r^2}$, Cu$_l$\,$4s$\,--\,O$_{b/a}$\,$2s'/2p'$,
O$_b^l$\,$2p'_y$--\,Cu$_l$\,$4p_y$--\,O$_b^c$\,$2p'_y$, and Cu$_l$
$3d'_{3z^2-r^2}$ character and occupation numbers of 1.999, 1.98,
0.01, 0.01, and 0.01, respectively. It can be seen that the nature
of some of these orbitals changes from the undistorted to the
distorted cluster. The CAS\,SCF results indicate that in the latter
case the O\,$2s/2p$\,--\,Cu\,$4s/4p$\,--\,O\,$2s/2p$ interactions
on the same plaquette are more important than the inter-plaquette
Cu\,$4s/4p$\,--\,O\,$2s/2p$\,--\,Cu\,$4s/4p$ interactions for the
former. Mulliken charges associated with such a ZR like state are
listed in Table\,V. The comparison between the results displayed
in Table\,V (or Table\,IV) and those obtained for the
electron-doped Cu$_l$O$_4$ plaquette in Table\,I reveals a
remarkable feature: due to strong charge redistribution effects
within the Cu$_l$ $3d_{3z^2-r^2}$, $4s$, $4p$ and O $2s$, $2p$
orbitals, the total electronic charge associated with the Cu$_l$
$3d_{x^2-y^2}^{\,1}$ ion in Table\,V (or Table\,IV) is nearly as
high as that associated with the Cu$_l$ $3d_{x^2-y^2}^{\,2}$
configuration from Table\,I, 28.7 versus 28.8.
It is well known that the Mulliken charges may be misleading in
cases where strong inter-atomic orbital overlap occurs. Still,
a variation of nearly one elementary charge within the Cu 
$4s$,$4p$ shell cannot be an artifact.

The adiabatic energy landscape for Cu--O$_b$ bond-length
distortions on the two central plaquettes is shown in Fig.4. The
points on the full lines are the minimal-CAS results already
published in Ref.\,\cite{cuo_HNY}. The energy maxima correspond to
states with dominant O$_b$-hole character in the undistorted
structure. The two minima are related to ZR like states where the
Cu--O$_b$ distances on the Cu$_l$O$_4$ plaquette (the left hand
minimum) or the Cu--O$_b$ distances on the Cu$_r$O$_4$ plaquette
(the right hand minimum) are $6\%$ shorter. From each
minimum-energy structural configuration, the two O$_b$ ligands are
shifted back to the undistorted geometry in steps of $1\%$ of the
high-symmetry Cu--O bond length. These states are shown with black
dots. The last point on each of the two curves corresponds to a
configuration where the Cu--O$_b$ distances on either the
Cu$_{ll}$O$_4$ or Cu$_{rr}$O$_4$ plaquette are shorter, by $1\%$.
Other calculations were carried out for the case where the
Cu--O$_b$ bonds on the Cu$_l$O$_4$ and Cu$_r$O$_4$ plaquettes are
gradually shortened starting from the undistorted structure and
using the orbitals of the O$_b^c$-hole state as starting orbitals.
Due to convergence problems, only few points could be obtained.
Those points are shown as black triangles.

Relative energies from CAS\,SCF calculations with 13 electrons and
12 orbitals in the active space are shown in Fig.4 only for the ZR
like states with Cu--O$_b$ distances shorter by $6\%$ and the
states with undistorted bonds. These points are marked with filled
squares. The energy lowering induced by the use of a larger active
space is of the order of 1.5--2.0\,eV. The effect is stronger for
the distorted configurations.
For convenience, the larger-CAS energies are shifted upwards in
Fig.4 by 1.6\,eV.
We note that the active orbitals and the CI vectors for the
states with dominant O$_b^l$ or O$_b^r$ -hole character are
similar to those of the state where the hole has the largest weight
on the O$_b^c$ ligand, with a ``shift'' of one plaquette to the left
or to the right, respectively.

\begin{figure}[!b]
\includegraphics[angle=0,width=1.00\columnwidth]{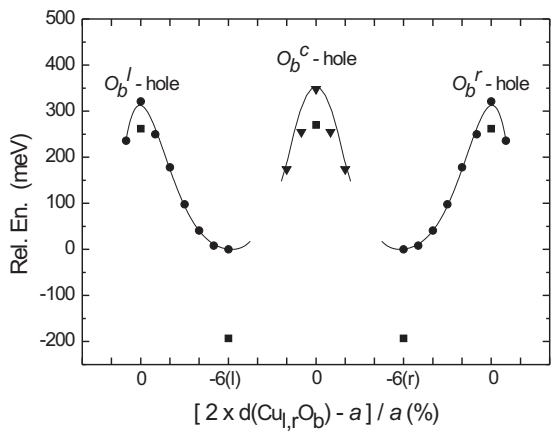}
\caption{ Results of static, total energy calculations for a
hole-doped, 4-plaquette cluster, see text.
Filled circles and triangles are the minimal-CAS
(\textit{i.e.}\,5\,electrons/5\,orbitals) results.
The lowest points are related to ZR like states where the
Cu$_l$--O$_b$ distances (the left hand minimum) or the Cu$_r$--O$_b$
distances (the right hand minimum) are $6\%$ shorter.
The energy maxima correspond to broken-symmetry states in the
undistorted geometry.
The filled squares are CAS\,SCF results with 13 electrons and 12
orbitals in the active space.
The minimal-CAS energy of the ZR state is taken as reference.
For convenience, the larger-CAS energies are shifted upwards by
1.6 eV. }
\end{figure}

Separately optimized, broken-symmetry wave-functions like the
different O-hole solutions for the undistorted geometry are
non-orthogonal and interacting. Orthogonal, non-interacting
eigenstates can be obtained by SI calculations. However, the
situation is now somewhat different as compared to the SI
calculations for the electron-doped cluster. First, the way the
energy landscape was drawn is different. The two sides of a
well in the hole-doped case correspond to identical positions of
the O$_b$ ions, although CAS\,SCF roots on different sides
have different character, at least for small distortions. In the
second place, the O-hole is ``smoothly'' transferred from one
plaquette to the other, for distortions of about 
$1\%$ of the Cu--O bond(s). Matrix elements exist thus only
between different O-hole states in the undistorted geometry or
between states situated on different sides of the same well.
We note that in the electron-doped cluster, matrix elements exist
only between CAS\,SCF states associated with different wells.
Those matrix
elements determine in the electron-doped case a so-called avoided
crossing for the energy curves. The missing points for each of the
two wells in Fig.4 could probably be obtained from calculations on
clusters including several 1-hole, 3-octahedra units. For certain
phases of the O-ion half-breathing vibrations on the neighboring
plaquettes, see Fig.1, the O-hole on a given 3-octahedra unit
should move ``smoothly'' from one side to the other. In fact,
it is the longer range Coulomb and spin interactions that can
induce indeed broken-symmetry oxygen-hole states in the
infinite system.
The collective, coherent ``tunneling'' of these quasi-localized
oxygen holes along a multi-well energy landscape like that depicted
in Fig.4 can lead to a traveling charge-density/spin-density wave as
in the model proposed by Goodenough \cite{jbg_03}.

The tendency towards the formation of quasi-localized oxygen-hole
states
on \textit{distorted} CuO$_4$ plaquettes is confirmed by model
Hamiltonian calculations. We consider in the following paragraphs
isolated chains of CuO$_4$ plaquettes at 1/3 hole-doping and
construct a $p\!-\!d$, Hubbard type model:
\begin{eqnarray*}
H &= &\epsilon_d \sum_{i\sigma} n_{i\sigma}
     +\epsilon_p \sum_{j\sigma} n_{j\sigma}
     +\sum_{<ij>\sigma}
           t_{pd}\,(d_{i\sigma}^{\dagger}p_{j\sigma} + h.c.)  \\
  &+ &\sum_{<jj'>\sigma}
           t_{pp}\,(p_{j\sigma}^{\dagger}p_{j'\sigma} + h.c.) \\
  &+ & U_{dd} \sum_i n_{i\uparrow}n_{i\downarrow}
     + U_{pp} \sum_j n_{j\uparrow}n_{j\downarrow}                 \\
  &+ & V_{pd} \sum_{<ij>\sigma\sigma'}  n_{i\sigma}n_{j\sigma'} +
       V_{pp} \sum_{<jj'>\sigma\sigma'} n_{j\sigma}n_{j'\sigma'} \ .
\end{eqnarray*}
We neglect the apical ligands and restrict to a single orbital at
each site: Cu $3d_{x^2-y^2\,}$, bridging oxygen O$_b$ $2p_y$
orbitals, and oxygen $2p_x$ orbitals at the other ligand sites,
O$_a$. We take into account the overlap integrals and Coulomb
interactions between the nearest neighbor $d_{x^2-y^2}$ and
$p_y$/$p_x$ $\sigma$-orbitals and also between the second nearest
neighbors O$_b$ $p_y$ and O$_a$ $p_x$. In the electron
representation, we choose
$\Delta\!=\!\epsilon_d\!-\!\epsilon_p\!=\!3.0$ \cite{oka_95} and
set $t_{pd}^x\!=-1.2$, $t_{pd}^y\!=-1.2$, and $t_{pp}\!=\!0.5$
(eV). The values of the Coulomb repulsion parameters are chosen
according to results of constrained LDA \cite{Upp_dft_88_89}
and/or \textit{ab initio}, explicitly correlated
\cite{cuo_stollhoff_98,cuo_RLmartin93,cuo_RLmartin96,cuo_calzado_02}
calculations:
$U_{dd}\!=\!9.0$, $U_{pp}\!=\!5.0$, $V_{pd}\!=\!1.5$, and
$V_{pp}\!=\!1.2$ (eV).
Under the assumption of charge segregation
into undoped and 1/3-doped stripes \cite{jbg_03}, we also include
for the O$_a$ ligands inter-site Coulomb interactions due to
electrons at Cu and O sites on the neighboring, undoped
Cu\,$d^9$--\,O\,$p^6$--... chains. That is, the energy of the
O$_a$ $p_x$ orbitals is shifted from
$\epsilon_p\!=\!\epsilon_p^y\!=\!\epsilon_d\!-\!3$ to
$\epsilon_p^x\!=\!\epsilon_{p}\!+\!V_{pd}\!+\!4\,V_{pp}$\,.

Ground-state expectation values are obtained by density-matrix
renormalization-group (DMRG) techniques \cite{dmrg_review_scholl}.
For these calculations chains of 33 CuO$_4$ plaquettes, open-end
boundary conditions, and $m\!=\!1200$ states to build the DMRG
basis were used. The typical discarded weight is of the order
$10^{-11}\!-\!10^{-10}$ or lower.

Occupation numbers of $p$ and $d$ orbitals on the 3-plaquette unit
at the middle of the 33-plaquette chain are listed in Table\,VI
(first column). The Cu site right at the middle is labeled
Cu$_c$. Along the Cu--O$_b$--... chain, the next sites to the left
and to the right are denoted O$_b^l$, Cu$_l$, O$_b^{ll}$ and
O$_b^r$, Cu$_r$, O$_b^{rr}$, respectively. For each of the Cu ions
mentioned above there is also a pair of O$_a$ sites. These ligands
are labeled O$_a^l$, O$_a^c$ or O$_a^r$. The occupation numbers
of the $d$ orbitals are comparable to those obtained in the
CAS\,SCF calculations for the small cluster. However, since the
O\,$2s,2p$\,$\rightarrow$\,Cu\,$4s,4p$ (back) charge transfer
mentioned above is not taken into account in our $p\!-\!d$ model
Hamiltonian, the electron occupation of the $p$ orbitals is
somewhat larger. Also, compared to the \textit{ab initio} results,
there is no charge disproportion along the undistorted chain of
CuO$_4$ plaquettes. This may be attributed to various effects. The
most important should be the neglect of longer range Coulomb
interactions in the $p\!-\!d$ model Hamiltonian and the orbital
optimization for the CAS\,SCF wave-function. The fact that the
variational optimization of the ``molecular'' orbitals can lead to
broken-symmetry solutions was shown for a simple model Hamiltonian
by Martin \cite{cuo_RLmartin96}.

\begin{table}[t]
\caption{Orbital occupation numbers on the 3-plaquette unit at the
middle of a chain of 33 CuO$_4$ plaquettes at 1/3 hole doping.
DMRG results for a $p\!-\!d$ extended Hubbard model, see text.
Three different geometries are considered. In the distorted
configurations, the Cu--O$_b$ bonds are shorter by $6\%$ on every
third plaquette. Each occupation number does not change by more
than 0.02 for the next four ``units''. Relative energies are also
given, for a whole chain.}
\begin{ruledtabular}
\begin{tabular}{lccc}
                   &Undist. &Dist. Cu$_c$--O$_b$
                            &Dist. Cu$_r$--O$_b$  \\
\colrule
\\
O$_b^{ll}$ $p_y$     &1.75    &1.80          &1.71      \\
Cu$_l$ $d_{x^2-y^2}$ &1.14    &1.18          &1.18      \\
O$_a^{l}$  $p_x$     &1.89    &1.93          &1.93      \\
O$_b^{l}$  $p_y$     &1.74    &\textbf{1.71} &1.81      \\
Cu$_c$ $d_{x^2-y^2}$ &1.13    &\textbf{1.04} &1.18      \\
O$_a^{c}$  $p_x$     &1.88    &\textbf{1.83} &1.93      \\
O$_b^{r}$  $p_y$     &1.74    &\textbf{1.71} &\textbf{1.71} \\
Cu$_r$ $d_{x^2-y^2}$ &1.14    &1.18          &\textbf{1.04} \\
O$_a^{r}$  $p_x$     &1.89    &1.93          &\textbf{1.83} \\
O$_b^{rr}$ $p_y$     &1.75    &1.80          &\textbf{1.71} \\
Rel. En. (eV)        &0       &--2.80 \      &--2.44 \      \\
\end{tabular}
\end{ruledtabular}
\end{table}

Next, we simulate the shortening of the Cu--O$_b$ bonds on every
third plaquette along the chain by introducing a modulation of the
hopping matrix elements and inter-site Coulomb repulsion
parameters: $t'_{pd}(r)\!=\!t_{pd}(0)\,(r_0/r)^{7/2}$,
$t'_{pp}(r)\!=\!t_{pp}(0)\,(r_0/r)^{3}$ \cite{book_harrison}, and
$V'(r)\!=\!V(0)\,(r_0/r)$. Orbital occupation numbers are given in
Table\,VI, for bonds shortened by $6\%$ on the central plaquette
on every 3-plaquette unit (second column) and for bonds shortened
by $6\%$ on the right-hand plaquette on every 3-plaquette unit
(third column). We also list the energies calculated for the three
configurations, with the energy of the undistorted chain taken as
reference. The results show that the energy of the system is
lowered for the distorted configurations. The small energy
difference between the two differently distorted structures, 2.80
versus 2.44 eV, is due to finite size effects.
In addition, the charge redistribution within the 3-plaquette unit(s)
confirms the trend observed for the CAS\,SCF results on the smaller
cluster: the oxygen holes tend to localize on the distorted
plaquettes.
Electron--phonon interactions and a softening of the half-breathing
O-ion vibrations were predicted in the hole-doped regime by other
authors as well \cite{tJ_and_phonons}, mostly on the basis of
$t\!-\!\!J$ model Hamiltonian studies.
With the present DMRG calculations for a $p\!-\!d$ extended Hubbard
model, we found that the energetic effects associated with
Cu--O$_b$ bond-length distortions on every third plaquette along a
chain of several plaquettes at 1/3-doping is of the order of
hundred meV per each 3-plaquette ``unit'' (approximately 255\,meV
for distortions of $6\%$). These estimates are in line with the
CAS\,SCF results discussed above. Nevertheless, the \textit{ab
initio} calculations indicate a much stronger effect.

\section{Conclusions}

Results of \textit{ab initio} multiconfiguration calculations on
clusters including few CuO$_4$ plaquettes and also results of DMRG
calculations for a $p\!-\!d$ extended Hubbard model with chains of
few tens of plaquettes indicate the presence of strong anharmonic
effects for the O-ion half-breathing deformations. In addition,
for certain phases of the O-ion vibrations, charge can be
displaced along the CuO chain. These effects are observed for both
hole and electron doping.

The conditions under which the CuO$_2$ layers can be doped were
discussed by Goodenough \textit{et al.}, see
Refs.\,\cite{jbg_SrNdCuO2_91,jbg_03} for example. Electrons are
not easily accepted if the in-plane lattice parameter is too
small, $a\!\lesssim\!3.93$\,\AA . On the other hand, holes can be
introduced if the lattice parameter is not larger than 3.87\,\AA .
Strong couplings between the Cu--O bond-length fluctuations and
the charge carriers, with an anharmonic, multi-well energy
landscape and strong charge transfer effects were also predicted
in Ref.\,\cite{jbg_03}. This prediction is confirmed by our
present study. Nevertheless, both the CAS\,SCF results on the
small clusters and the DMRG estimates for the $p\!-\!d$, Hubbard
like CuO$_4$--\,CuO$_4$--... chain are rather qualitative. For the
CAS\,SCF calculations, a large part of the dynamic electron
correlation effects is not accounted for \cite{dyn_corr}. Coulomb
and (super)exchange interactions beyond the 4-plaquette cluster are
also neglected. Within the hole-doped, $p\!-\!d$ extended Hubbard
model we have neglected ``rehybridization'' effects
\cite{cuo_RLmartin96} arising from the variational optimization of
the orbitals.
Charge disproportion along the CuO chain can be induced then only by
longer range Coulomb interactions and structural distortions.
However, since interactions beyond the nearest neighbors are not
explicitly included, longer range Coulomb ``correlations'' occur
through indirect, higher order effects.
Each of the two approaches has thus its limitations, but to some
extent the two types of methods are complementary.
Most importantly, they both predict that the electron\,--\,lattice
couplings in the doped CuO$_2$ planes are strong and anomalous.
The superconductivity might be related to the collective,
coherent tunneling of the doped carriers along the multi-well energy
landscape.
Tunneling effects associated with an electron hole in a double-well
potential were analyzed on an oversimplified, hole-doped
[CuO$_{2}$]$^{-}$ cluster model by Bishop \textit{et al.}, see
\cite{CuO_bishop03_tu} and references therein.
The height of the energy barrier was 400\,meV in their model,
similar to our results. The amplitude of the O-ion displacements
was about two times smaller.

Within the traveling CDW/SDW scenario illustrated schematically in
Fig.1, the rigidity of the superconducting condensate is ensured by
the inter-carrier Coulomb repulsion and the relatively large AFM
interactions between the intervening Cu $d^9$ ions.
For a single chain, these couplings are weakened for doping levels
below 1/3.
In the case of hole doping, it seems that at higher concentrations
an homogeneous metallic state is realized, like the 1/2-doped
metallic chains in La$_{1.48}$Nd$_{0.40}$Sr$_{0.12}$CuO$_4$
\cite{stripes_tranquada95}.
The charge segregation into 1/3-doped and undoped stripes at optimal
doping \cite{jbg_03} is supported by inelastic neutron scattering
data \cite{mcqueeney_99,egami_00}.
Nevertheless, the existence of this type of charge ordering in the
CuO$_2$ plane is not definitely established.

\

We thank A. Yamasaki, G. Stollhoff, G. Kalosakas, and A. T. Filip
for fruitful discussions.
L.\,H. acknowledges partial financial support from the Alexander
von Humboldt Foundation.


\begin{thebibliography}{9}

\bibitem{cuo_stollhoff_91} C.-J. Mei and G. Stollhoff, Phys. Rev. B
{\bf 43}, 3065 (1991).

\bibitem{cuo_stollhoff_98} G. Stollhoff, Phys. Rev. B {\bf 58},
9826 (1998).

\bibitem{cuo_RLmartin93} R. L. Martin, J. Chem. Phys. {\bf 98},
8691 (1993).

\bibitem{cuo_RLmartin96} R. L. Martin, Phys. Rev. B {\bf 53},
15501 (1996).

\bibitem{cuo_calzado_00} C. J. Calzado, J. F. Sanz, and J. P.
Malrieu, J. Chem. Phys. {\bf 112}, 5158 (2000).

\bibitem{cuo_HNY} L. Hozoi, S. Nishimoto, and A. Yamasaki,
Phys. Rev. B {\bf 72}, 144510 (2005).

\bibitem{ZR_86} F. C. Zhang and T. M. Rice, Phys. Rev. B {\bf 37},
3759 (1988).

\bibitem{cuo_SM_92} \textit{Phase Separation in Cuprate
Superconductors}, Eds. K. A. M\"{u}ller and G. Benedek
(World Scientific, Singapore, 1992).

\bibitem{cuo_SM_94} \textit{Phase Separation in Cuprate
Superconductors}, Eds. E. Sigmund and K. A. M\"{u}ller
(Springer-Verlag, Berlin, 1994).

\bibitem{kivelson_stripes03} S. A. Kivelson, I. P. Bindloss,
E. Fradkin, V. Oganesyan, J. M. Tranquada, A. Kapitulnik, and
C. Howald, Rev. Mod. Phys. {\bf 75}, 1201 (2003).

\bibitem{CuO_bishop03_tu} A. R. Bishop, D. Mihailovic, and J.
Mustre de Le\'{o}n, J. Phys.: Condens. Matter {\bf 15}, L169 (2003).

\bibitem{jbg_03} J. B. Goodenough, J. Phys.: Condens. Matter
{\bf 15}, R257 (2003).

\bibitem{cuo_hizhny_88} V. Hizhnyakov and E. Sigmund, Physica C
{\bf 156}, 655 (1988).

\bibitem{cuo_zaanen_88} J. Zaanen and A. M. Ole\'{s}, Phys. Rev. B
{\bf 37}, 9423 (1988).

\bibitem{cuo_emery_88} V. J. Emery and G. Reiter, Phys. Rev. B
{\bf 38}, 4547 (1988).

\bibitem{CuO_Kochelaev97} B. I. Kochelaev, J. Sichelschmidt, B.
Elschner, W. Lemor, and A. Loidl, Phys. Rev. Lett. {\bf 79}, 4274
(1997).

\bibitem{QS_99_00} L. Proville and S. Aubry, Eur. Phys. J. B {\bf 15}, 
405 (2000); L. Proville and S. Aubry, Eur. Phys. J. B {\bf 11}, 41
(1999); see also S. Aubry in \cite{cuo_SM_92}.

\bibitem{alexandrov_95_03} A. S. Alexandrov and N. Mott,
\textit{Polarons and Bipolarons} (World Scientific, Singapore, 1995);
A. S. Alexandrov, \textit{Theory of Superconductivity: From Weak
to Strong Coupling} (Taylor and Francis, London, 2003).

\bibitem{NaVO_phi4} L. Hozoi, C. Presura, C. de Graaf, and R. Broer,
Phys. Rev. B {\bf 67}, 035117 (2003); L. Hozoi, S. Nishimoto, and A.
Yamasaki, Phys. Rev. B {\bf 72}, 195117 (2005).

\bibitem{book_qc} For a monograph, see T. Helgaker, P. J\o rgensen,
and J. Olsen, \textit{Molecular Electronic-Structure Theory}
(Wiley, Chichester, 2000).

\bibitem{jbg_SrNdCuO2_91} M. G. Smith, A. Manthiram, J. Zhou,
J. B. Goodenough, and J. T. Markert, Nature {\bf 351}, 549 (1991).

\bibitem{ANOs_CuO} R. Pou-Am\'{e}rigo, M. Merch\'{a}n, I. Nebot-Gil,
P.-O. Widmark, and B. O. Roos, Theor. Chim. Acta {\bf 92},
149 (1995); P.-O. Widmark, P.-\AA. Malmqvist, and B. O. Roos,
Theor. Chim. Acta {\bf 77}, 291 (1990).

\bibitem{cuo_munoz_02} D. Mu\~{n}oz, I. de P. R. Moreira, and F. Illas,
Phys. Rev. B {\bf 65}, 224521 (2002).

\bibitem{cuo_calzado_02} C. J. Calzado, J. Cabrero, J. P. Malrieu,
and R. Caballol, J. Chem. Phys. {\bf 116} 3985 (2002). 

\bibitem{coen_thesis} C. de Graaf, Ph. D. thesis, Rijksuniversiteit
Groningen (1998).

\bibitem{TIPs_CuSr} F. Illas, J. Rubio, J. C. Barthelat,
Chem. Phys. Lett. {\bf 119}, 397 (1985);
W. R. Wadt and P. J. Hay, J. Chem. Phys. {\bf 82}, 284 (1985).

\bibitem{molcas6} \textsc{molcas 6}, Department of Theoretical
Chemistry, University of Lund, Sweden.

\bibitem{dd_77_81} B. H. Botch, T. H. Dunning, and J. F. Harrison, 
J. Chem. Phys. {\bf 75}, 3466 (1981); C. Froese Fischer, J. Phys. B
{\bf 10}, 1241 (1977).

\bibitem{oka_95} O. K. Andersen, A. I. Liechtenstein, O. Jepsen,
and F. Paulsen, J. Phys. Chem. Solids {\bf 56}, 1573 (1995).

\bibitem{oka_01} E. Pavarini, I. Dasgupta, T. Saha-Dasgupta,
O. Jepsen, and O. K. Andersen, Phys. Rev. Lett. {\bf 87}, 047003
(2001).

\bibitem{dd_92} K. Andersson and B. O. Roos, Chem. Phys. Lett.
{\bf 191}, 507 (1992).

\bibitem{ppdd_coen_01} C. de Graaf, C. Sousa, I. de P. R. Moreira,
and F. Illas, J. Phys. Chem. A {\bf 105}, 11371 (2001).

\bibitem{SI_86_89} P.-\AA. Malmqvist, Int. J. Quantum. Chem.
{\bf 30}, 479 (1986); P.-\AA. Malmqvist and B. O. Roos,
Chem. Phys. Lett. {\bf 155}, 189 (1989).

\bibitem{CuO_eldo_sun04} X. F. Sun, Y. Kurita, T. Suzuki, S.
Komiya, and Y. Ando, Phys. Rev. Lett. {\bf 92}, 047001 (2004).

\bibitem{LaCuO_cava87} R. J. Cava, A. Santoro, D. W. Johnson, and
W. W. Rhodes, Phys. Rev. B {\bf 35}, 6716 (1987).

\bibitem{note_localization} Compared to the minimal-CAS results
reported in Ref.\,\cite{cuo_HNY}, the present study shows
that the tendency towards localization on a single anion is less
pronounced for larger active spaces.
However, we continue to call these broken-symmetry states in the
undistorted structure as O$_b^l$, O$_b^c$ or O$_b^r$ -hole states.

\bibitem{geopt_mno_nio} M. D. Towler, N. L. Allan, N. M. Harrison,
V. R. Saunders, W. C. Mackrodt, and E. Apr\'{a}, Phys. Rev. B
{\bf 50}, 5041 (1994).

\bibitem{geopt_cu2o} E. Ruiz, S. Alvarez, P. Alemany, and
R. A. Evarestov, Phys. Rev. B {\bf 56}, 7189 (1997).

\bibitem{note_HFgo} Geometry optimizations at a correlated, post-HF
level are not possible yet.

\bibitem{Upp_dft_88_89} A. K. McMahan, R. M. Martin, and
S. Satpathy, Phys. Rev. B {\bf 38}, 6650 (1988);
M. S. Hybertsen, M. Schl\"{u}ter, and N. E. Christensen,
Phys. Rev. B {\bf 39}, 9028 (1989).

\bibitem{dmrg_review_scholl} For a review, see U. Schollw\"{o}ck,
Rev. Mod. Phys. {\bf 77}, 259 (2005).

\bibitem{book_harrison} W. A. Harrison, \textit{Electronic Structure
and the Properties of Solids} (Dover, New York, 1989).

\bibitem{tJ_and_phonons} See for example
\cite{CuO_szczepanski96,CuO_oliver04,CuO_ishihara04,CuO_piekarz05}
and references therein.

\bibitem{CuO_szczepanski96} K. J. von Szczepanski and K. W. Becker,
Z. Phys. B {\bf 89}, 327 (1992).

\bibitem{CuO_oliver04} O. R\"{o}sch and O. Gunnarsson, Phys. Rev.
Lett. {\bf 92}, 146403 (2004).

\bibitem{CuO_ishihara04} S. Ishihara and N. Nagaosa, Phys. Rev. B
{\bf 69}, 144520 (2004).

\bibitem{CuO_piekarz05} P. Piekarz and T. Egami, Phys. Rev. B
{\bf 72}, 054530 (2005).

\bibitem{dyn_corr} More sophisticated methods, e.g., multi-reference
CI or multiconfiguration perturbation theory, are needed for
treating dynamical correlation effects \cite{book_qc}. Such
calculations are beyond the scope of this paper.

\bibitem{stripes_tranquada95} J. M. Tranquada, B. J. Sternlieb,
J. D. Axe, Y. Nakamura, and S. Uchida, Nature {\bf 375}, 561 (1995).

\bibitem{mcqueeney_99} R. J. McQueeney, Y. Petrov, T. Egami,
M. Yethiraj, G. Shirane, and Y. Endoh, Phys. Rev. Lett. {\bf 82},
628 (1999).

\bibitem{egami_00} T. Egami, Y. Petrov, and D. Louca, J. Supercond.:
Incorp. Novel Magn. {\bf 13}, 709 (2000);
T. Egami, J.-H. Chung, R. J. McQueeney, M. Yethiraj, H. A. Mook,
C. Frost, Y. Petrov, F. Dogan, Y. Inamura, M. Arai, S. Tajima, and
Y. Endoh, Physica B {\bf 316--317}, 62 (2002).


\end{thebibliography}
\end{document}